\documentstyle[preprint,amstex,aps,prb]{revtex}
\def\ds#1{\renewcommand{\baselinestretch}{#1}}
\ds{1.5}
\begin{title}
{ The Puzzling Collapse of Electronic Sliding Friction on a Superconductor
Surface}
\end{title}

\author{B.N.J. Persson$^1$ and E. Tosatti$^2$\\
$^1$ {\it Institut f{\"u}r Festk{\"o}rperforschung,
Forschungszentrum J{\"u}lich,
D--52425 J{\"u}lich, Germany}\\
$^2${\it International School for Advanced Studies (SISSA),
Via Beirut 2, I-34014 Trieste, Italy, and \\
Istituto Nazionale di Fisica della Materia (INFM), Trieste,
(Italy), and \\
Abdus Salam International Center for Theoretical Physics (ASICTP),
P.O. Box 586, I-34014 Trieste, Italy}}

\begin{document}

\maketitle

ABSTRACT\\
In a recent paper [Phys. Rev. Lett. {\bf 80} (1998) 1690],
Krim and coworkers have observed that the friction force, acting
on a thin physisorbed layer of N$_2$ sliding on a lead film, abruptly
decreases by a factor of $\sim 2$ when the lead film is cooled
below its superconductivity transition temperature. We discuss the
possible mechanisms for the abruptness of the sliding friction
drop, and also discuss the relevance of these results to the
problem of electronic friction.

\vskip 1cm

The frictional forces and dissipation experienced by thin physisorbed
layers of inert atoms or molecules, such as Kr, Xe or N$_2$  while
sliding on metallic surfaces can nowadays be measured, owing to an
ingenious method devised by J. Krim and collaborators.\cite{Krim1}

The frequency shift and, most importantly, the change in the $Q$-factor
experienced by a quartz crystal microbalance
(oscillating at about $\sim 10 {\rm MHz}$),
upon adsorption of the gas on the metal film, itself part of the
microbalance, provide direct information on the molecule-surface
frictional processes.

In this manner both the phononic and electronic contribution to friction
can be accessed. In particular, when the substrate is a metal that can
be cooled down below the superconducting $T_c$, one can gauge the
importance of the electronic contribution, as this alone should
presumably change below $T_c$. This is precisely what was done in a very
recent experiment,\cite{Krim2} where about 1.6 ML of N$_2$ was adsorbed
on a lead film, which can be cooled below the lead film superconducting
temperature $T_c \approx  7{\rm K}$.

The result is striking: dissipation due to friction drops at $T_c$
to about half of its normal state value. The drop is clearly
connected with superconductivity of the metal substrate, and is
very abrupt. While the phenomenon confirms predictions about the importance
of electronic friction (see below), its abruptness is
entirely unexpected, and has not as yet been explained or even discussed
to any level of detail. The purpose of this note is to debate on
the possible mechanisms for the abruptness of the sliding friction drop,
and also to discuss the relevance of these results to the
problem of electronic friction.

At first one may think that the
explanation for the abruptness is trivial.
When the metal film is in the normal state, at $T>T_c$, it is
known that  the electronic contribution to the sliding friction
is directly proportional to the resistivity change of the metal film
induced by the adsorbate layer.\cite{Persson2}
This is proven by comparing
the energy ``dissipation'' calculated in the reference frame where
the metal film is stationary and the adsorbate layer moves
with the velocity ${\bf v}$ (``frictional energy dissipation''),
with that calculated in the frame where the adsorbate
layer is stationary and the metal electrons move with
velocity ${\bf -v}$ (``ohmic energy dissipation'').\cite{Persson2}
If this argument were to remain valid also when the metal film is
in the superconducting state, then the electronic contribution to the sliding
friction would correctly drop abruptly to zero at $T=T_c$, since
the film resistivity vanishes abruptly at $T_c$.

We argue however that this argument is incorrect, unfortunately,
when the metal film is in the superconducting state. First we note that the
superconductivity transition is continuous, so that the
fraction of the electrons in the superconducting condensate increases
{\it continuously} from zero to one as the temperature is reduced
from  $T_c$ to zero. The DC
resistivity of the metal film drops nonetheless discontinuously
from its normal state value above $T_c$ to zero at $T=T_c$. The reason
is, of course, that the electrons in the condensate short
circuit the metal film. Thus, no drift motion (current) occurs
in the ``normal'' electron fluid of thermal  excitations,
even though just below $T_c$ almost all the electrons belong in
this ``normal'' fluid.  Let us now consider the
system in a reference frame where the adsorbate
layer is stationary.

In this reference frame all the electrons
in the metal film move (collectively) with the speed ${\bf -v}$
relative to the adsorbate layer. The ``normal'' electrons in the system
will scatter from the adsorbates and give rise to energy dissipation,
just like in the normal state (i.e., $T>T_c$).
This is precisely what happens, for example, in ultrasonic attenuation.
\cite{Schrieffer} By analogy, one would thus expect the electronic
sliding friction to decrease {\it continuously} as the
system is cooled below $T_c$, in a way which correlates with the
fraction of electrons in the condensate, and in sharp contrast to what is
observed experimentally.

Of course, the perturbation represented by the sliding adsorbate film
is localized at the surface, rather than extended as in the case of
ultrasonic attenuation. The localized processes should necessarily
involve $k_z$-nonconservation, corresponding to a strong spatial variation,
which is not present in bulk ultrasonic attenuation (but which might be
achievable in {\it surface} ultrasonic attenuation). There is
however no immediate reasoning suggesting that precisely this fact
should cause the switching from continuous to abrupt.

Again by analogy to ultrasonic attenuation, a jump in the electronic
friction at $T=T_c$ could still be explained, if a transverse
electromagnetic field were somehow involved in the excitation process.
The superconducting condensate (even if very weak, near $T_c$),
will screen out the transverse electromagnetic field in the
metal and hence abruptly eliminate the coupling between the normal
electrons and the adsorbate. This effect is well
known and is, e.g., the reason why the damping of transverse acoustic
phonons jumps abruptly, by roughly a factor of two, at
$T=T_c$.\cite{Schrieffer}

On the other hand, existing understanding of the electronic sliding friction
in the normal state \cite{Persson3}  suggests that the short-range,
unretarded coupling between the
fluctuating coulomb field associated with virtual electronic
transitions in the adsorbates should be quite adequate. Within the short depth
below the surface where most of the sliding-induced electron-hole
excitations in the metal are generated, this fluctuating
field can be considered as longitudinal. Consequently one would
not expect an abrupt drop in the electronic friction due to this
coupling mechanism.

The $N_2$ layer adsorbed on lead at this low temperature must be
essentially solid, probably incommensurate. The friction jump
could in principle signal a supersolid behavior of the film.
However, again we see no reason why this should happen, and why precisely
at the substrate superconducting $T_c$.

The lead surface employed by Krim was not a clean one, but had been
exposed for some time to atmospheric contamination. It could therefore
be covered by a thin layer of nonmetallic material, including oxide.
Friction on a nonmetal, particularly if polar, may give rise to
time-varying electric fields (triboelectricity), at least partly
tranverse, which in turn could be dissipated in the underlying metal
and be responsible for the observed jump. At present however this
kind of possibility is purely speculative and unsupported by any
known fact.

The abruptly jumping friction remains therefore a puzzle, to which
we wish to call attention, and for which we can offer at present
no clear-cut explanation.

The new phenomenon\cite{Krim2}, if confirmed, definitely
proves the importance of the electronic friction in the sliding of
incommensurate adsorbate layers. In the present system, at least half of the
sliding friction in the normal state must be of electronic origin.
This result is all the more remarkable since the evaporated lead film
is certainly not perfect and crystalline, and moreover was exposed
to air for 10-15 minutes. The resulting oxide layer should have
reduced the coupling between the sliding molecular film and the metal
conduction electrons,  while enhancing the coupling to phononic
excitations in the sliding film, in virtue of a larger atomic
corrugation than that of a clean metal surface.

These results and arguments suggest that on a clean well-defined metal
surface, e.g., Xe on  clean silver and gold films,\cite{Krim3} electronic
friction should be even more important than in Krim's latest experiment,
possibly dominating the whole sliding friction of incommensurate
adsorbate layers. This conclusion
is consistent with ideas previously put forward
by one of us\cite{Persson1} but not with those of other
authors.\cite{others}\\

\vskip 1cm

ACKNOWLEDGMENTS

We thank J. Krim and Ch. W\"oll for interesting discussions.
Work at SISSA is partly sponsored by INFM under PRA LOTUS, and by MURST
under contract COFIN97.

\end{document}